\documentclass[12pt,preprint]{aastex}

\newcommand{\cmpss}{cm~s$^{-2}$}
\newcommand{\mps}{m~s$^{-1}$}
\newcommand{\kps}{km~s$^{-1}$}
\newcommand{\Msun}{${\rm M_\odot}$}
\newcommand{\Rsun}{${\rm R_\odot}$}
\newcommand{\Mjup}{${\rm M_J}$}
\newcommand{\Rjup}{${\rm R_J}$}
\newcommand{\vMs}{1.0}		
\newcommand{\eMs}{0.03}
\newcommand{\vRs}{1.0}		
\newcommand{\eRs}{0.08}
\newcommand{\sptype}{G1V}
\newcommand{\vrvK}{116}		
\newcommand{\ervK}{9}
\newcommand{\vDs}{200}		
\newcommand{\eDs}{20}
\newcommand{\vjd}{2453808.9170}	
\newcommand{\ejd}{0.0011}	
\newcommand{\vap}{0.0488}	
\newcommand{\eap}{0.0005}
\newcommand{\vperiod}{3.941534}	
\newcommand{\eperiod}{0.000027}
\newcommand{\vMp}{0.90}		
\newcommand{\eMp}{0.07}	
\newcommand{\vRp}{1.30}		
\newcommand{\eRp}{0.11}
\newcommand{\vincl}{87.7}		
\newcommand{\eincl}{1.2}
\newcommand{\valq}{0.0275}		
\newcommand{\errq}{0.0005}

\slugcomment{To appear 2006.09.01 in the Astrophysical Journal}

\received{2006 Apr 06}
\begin{document}

\title{A Transiting Planet of a Sun-like Star}

\author{
P.~R.~McCullough\altaffilmark{1,2},
J.~E.~Stys\altaffilmark{1},
Jeff~A.~Valenti\altaffilmark{1},
C.~M.~Johns-Krull\altaffilmark{3,4},
K.~A.~Janes\altaffilmark{5}, 
J.~N.~Heasley\altaffilmark{6},
B.~A.~Bye\altaffilmark{2},
C.~Dodd\altaffilmark{2},
S.~W.~Fleming\altaffilmark{7},
A.~Pinnick\altaffilmark{5},
R.~Bissinger\altaffilmark{8},
B.~L.~Gary\altaffilmark{9},
P.~J.~Howell\altaffilmark{5},
T.~Vanmunster\altaffilmark{10}
}

\email{pmcc@stsci.edu}

\altaffiltext{1}{Space Telescope Science Institute, 3700 San Martin Dr., Baltimore MD 21218}
\altaffiltext{2}{University of Illinois, Urbana, IL 61801}
\altaffiltext{3}{Dept. of Physics and Astronomy, Rice University, 6100 Main Street, MS-108, Houston, TX 77005}
\altaffiltext{4}{Visiting Astronomer, McDonald Observatory, which is operated by the University of Texas at Austin.}
\altaffiltext{5}{Boston University, Astronomy Dept., 725 Commonwealth Ave.,
Boston, MA 02215}
\altaffiltext{6}{University of Hawaii, Inst. for Astronomy, 2680 Woodlawn Dr., Honolulu, HI 96822-1839}
\altaffiltext{7}{Dept. of Astronomy, University of Florida, 211 Bryant Space Science Center, Gainesville, FL 32611}
\altaffiltext{8}{Racoon Run Observatory, Pleasanton, CA}
\altaffiltext{9}{Hereford Arizona Observatory, 5320 E. Calle Manzana, Hereford, AZ 85615}
\altaffiltext{10}{CBA Belgium Observatory, Walhostraat 1A, B-3401 Landen, Belgium}

\begin{abstract}
A planet transits an 11$^{\rm th}$ magnitude, \sptype\ star in the
constellation Corona Borealis. 
We designate the planet XO-1b, and the star, XO-1, also known as 
GSC 02041-01657. 
XO-1 lacks a trigonometric distance; we estimate
it to be \vDs$\pm$\eDs\ pc.
Of the ten stars currently known to host extrasolar transiting planets,
the star XO-1 is the most similar to the Sun in its physical characteristics:
its radius is \vRs$\pm$\eRs\ \Rsun, 
its mass is \vMs$\pm$\eMs\ \Msun, V~sin$i < 3$\ \kps, and its metallicity
[Fe/H] is 0.015$\pm$0.04.
The orbital period of the planet XO-1b is \vperiod$\pm$\eperiod\ days,
one of the longer ones known.
The planetary mass is \vMp$\pm$\eMp\ \Mjup,
which is marginally larger than that of other transiting planets with
periods between 3 and 4 days.
Both the planetary radius and the inclination
are functions of the spectroscopically determined stellar radius.
If the stellar radius is \vRs $\pm$\eRs\ \Rsun, then the planetary radius
is \vRp$\pm$\eRp\ \Rjup\ and the inclination of the orbit
is \vincl$\pm$\eincl\arcdeg. 
We have demonstrated a productive international
collaboration between professional and amateur astronomers that was important to
distinguishing this planet from many other similar candidates.
\end{abstract}


\section{Introduction}

We know of five planets that transit stars brighter than V=12: HD209458b,
TrES-1b, HD 149026b, HD 189733b, and (now) XO-1b (Charbonneau et al. 2000
and Henry et al. 2000; Alonso et al. 2004; Sato et al. 2005; Bouchy et al. 2005a, and this work,
respectively report the discoveries).\footnote{Alonso et
al (2004) referred to both the planet and its host star as ``TrES-1,''
but for clarity we prefer the trailing ``b'' for the planets.} 
Observations of planetary transits of fainter stars still permit a rather
precise 
measure of the relative size of a star and its planet, but the brighter
stars provide adequate flux to enable interesting follow-up observations,
including observations of atmospheric composition
(Charbonneau et al.\ 2002; Vidal-Madjar et al.\ 2004), exosphere extent
(Vidal-Madjar et al.\ 2003), and infrared emission (Charbonneau et al.\
2005; Deming et al.\ 2005). 
The five remaining systems, discovered by OGLE (e.g. Bouchy et al. 2005b),
are too faint (V$\sim$15) for such studies. 
The XO project aims to find planets transiting
stars sufficiently bright to enable interesting follow up
studies (McCullough et al. 2005).\footnote{This paper includes data taken
on the Haleakala summit maintained by the University of Hawaii,
the Lowell Observatory,
the McDonald Observatory of The University of Texas
at Austin, and four backyard observatories.}

In addition to atmospheric studies of planets, transits around bright
stars also provide the best constraints on planetary interior models,
which currently cannot explain the observations (Konacki et al.\ 2005).
They also can provide an empirical determination of stellar
limb darkening and the frequency and contrast of star spots (Silva 2003).
Transits provide simple 1-dimensional imaging of the stellar photospheres,
because an inverted transit light curve represents a 1-dimensional
trace through the 2-dimensional convolution of the star's brightness
distribution with a uniform disk equal in radius to the planet.  In an
analogous manner, eclipses in principle can provide the brightness
distribution on the face of the planet as the star acts as a knife edge
cutting across the disk of the planet during ingress or egress.
Herein, we use ``transit'' to describe a planet passing
in front of a star, and ``eclipse'' to describe a star passing in front of
a planet or another star. 

Comparison studies will become increasingly interesting and important as
different methods of discovering transiting extrasolar planets produce additional
systems. Some radial-velocity surveys (Robinson et al. 2006; da Silva et
al. 2006) have a selection bias for high metallicity stars, in order to
increase the fraction of observed stars that have hot Jupiters (Valenti
\& Fischer 2005).  Photometric surveys for transits are biased toward
large depths of the transits (i.e large planets and/or small stars)
and short periods, both of which increase detectability.

One of the challenges of transit surveys such as the XO project is
distinguishing the stars with transiting planets from those without. Initial
discovery of candidate planets requires observing thousands of stars,
many hundreds of times each, with $\sim$1\%\ or better photometry per
epoch.  A light curve consistent with a transiting planet more
often is due to a grazing eclipsing binary star, a triple star system,
or a stellar companion that is of planetary size but not of planetary
mass (Brown 2003).  

Whereas high-resolution spectroscopy is required to measure a planetary
companion's mass, either
moderate-resolution spectroscopy or high-fidelity photometry in many cases can
distinguish the grazing eclipsing binary stars and the triple stars
(Section \ref{sec:triple}) from transiting planets.
Practical considerations may determine which is the preferred method
(spectroscopy or photometry).
A spectrum will often reveal a double-lined eclipsing
binary immediately, and a pair of spectra separated by mere hours
can reveal the large amplitude (\kps) radial velocity
variation of a single-lined eclipsing binary (or multiple) star system.
Such spectra can be obtained with moderate aperture (e.g. 2-m diameter)
telescopes and with little regard for the ephemeris of the transits.
On the other hand, precise photometry of transits demands that the
candidate be observed at specific and predictable times that nevertheless
are inconveniently infrequent for observing in a ``traditional'' manner
by an astronomer who must wait months to observe in hopes that during
the week or two that he or she is allocated on a telescope, and the
sky is clear, and the sun is down, and the star is at low to moderate
zenith angle, the planet will transit. Faced with such prospects, some
astronomers have deployed dedicated robotic telescopes for photometric
follow up (e.g. Sleuth of O'Donovan, Charbonneau, \& Kotredes 2004; TopHat of
Bakos et al. 2006a).

We chose a different tactic: we formed a collaboration of professional
and amateur astronomers. The high-quality photometry produced by
amateur astronomers of the transits of HD 209458 and Tres-1 (Naeye 2004)
immediately after their announcements by professional astronomers
proved that amateur astronomers can produce photometry of individual
stars that exceeds the quality obtainable by an instrument (such
as XO) that was designed to survey thousands of stars quickly. The
XO P.I. recruited four amateur astronomers (T. V., R. B., B. G., and
P. H.) to assist with photometric follow up of candidates identified
in the XO data. Important criteria in selecting these individuals were
1) the observatories should be distributed in longitude,
permitting a given transit to have a high probability of being observed
quickly, 2) each member should have exclusive access to a CCD-equipped
telescope and the expertise and software to
produce differentially-calibrated time-series photometry,
3) the telescope(s) should be capable of unattended operation,
in order to reduce attrition even
though only a small fraction of targets would be genuine planets, and 4)
internet access and a team-oriented attitude.  Henden (2006) describes
other amateur-professional collaborations; one related to transiting
planets is transitsearch.org (Seagroves et al. 2003).

Whereas some have advocated dispersing survey instruments longitudinally
around the globe (e.g. Alonso et al. 2004), the example of XO-1b
demonstrates the effectiveness of a survey instrument such as XO on a
single excellent site such as Haleakala, especially if it is coupled to a
longitudinally dispersed network of observatories capable of subsequent
study of candidates individually at specific times.  We hope that as
the XO Extended Team (E.T.) of advanced amateur astronomers grows in
members, its capabilities and its value to extrasolar planet discovery will grow
geometrically.

\section{Observations}

\subsection{XO Project Photometry}

\begin{figure}
\plotone{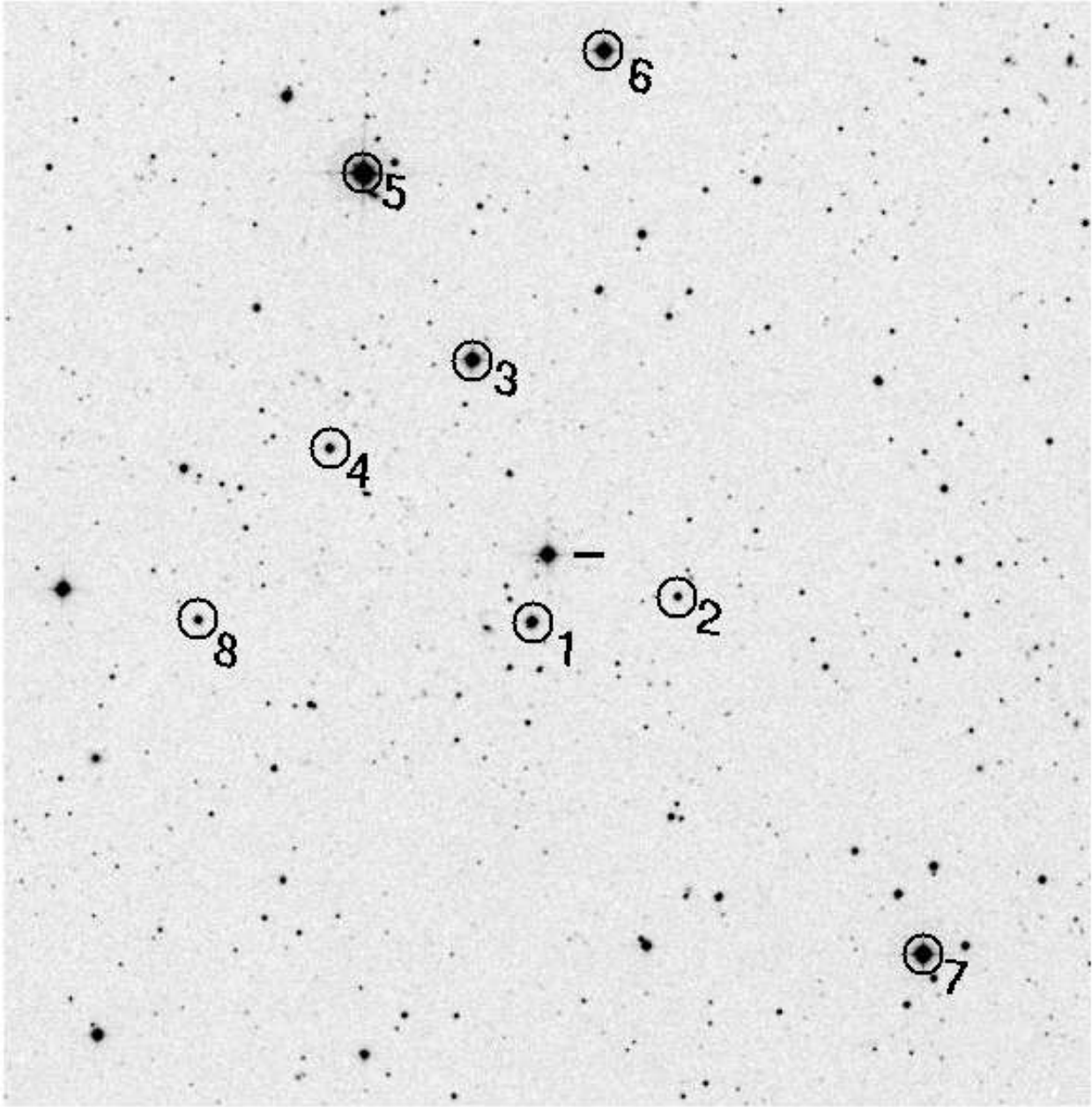}
\caption{XO-1 is centered, indicated by the hash mark.
Stars from Table \ref{table:allsky} are circled. 
North is up; East to the left. The DSS image, digitized from a
POSSII-F plate with a IIIaF emulsion and an RG610 filter,
subtends 19\arcmin\ of declination.
\label{fig:allsky}}
\end{figure}

\begin{figure}
\plotone{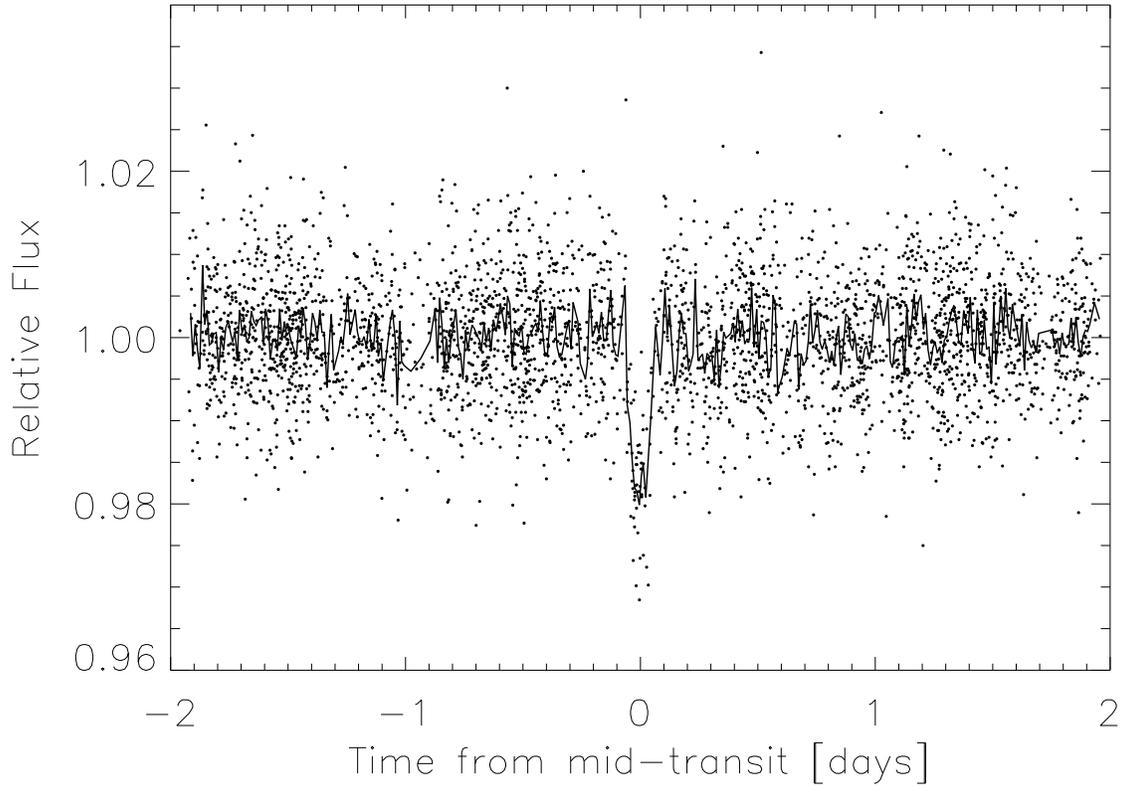}
\caption{More than 3000 individual observations of XO-1 by the two XO cameras
over two seasons 2004 and 2005 are shown wrapped and phased according
to the ephemeris of Equation \protect{\ref{eq:ephem}} and averaged
in 0.01-day bins (line).
From these data we identified the star
as a candidate for more refined photometry with other telescopes at
epochs of expected transits (Figure \ref{fig:etlc}).
\label{fig:xolc}}
\end{figure}

McCullough et al. (2005) describe the instrumentation, operation,
analysis, and preliminary results of the XO project.\footnote{The analysis
of XO photometry is incomplete, so a statistical
analysis would be premature and would tend to underestimate the efficiency
for discovering similar transiting planets within the data.}
In summary, the XO
observatory monitored tens of thousands of bright (V$<12$) stars twice
every ten minutes on clear nights for more than 2 months per season
of visibility for each particular star, over the period September 2003
to September 2005. From our analysis of more than 3000 observations per
star, we identified XO-1 (Figure \ref{fig:allsky}) as one of dozens of stars
with light curves suggestive of a transiting planet. 
With the XO cameras on Haleakala, we observed three transits of XO-1
in 2004 and one in 2005, on Julian dates 2453123, 2453127, 2453143, and 2453454.
From the survey photometry of XO-1 (Figure \ref{fig:xolc}),
which has a nominal standard deviation of 0.8\% or 8 mmag
per observation, we determined a preliminary light
curve and ephemeris, which we used to schedule
observations of higher quality with other telescopes,
as described in the next section. 

\subsection{Additional Photometry}

\begin{figure}
\plotone{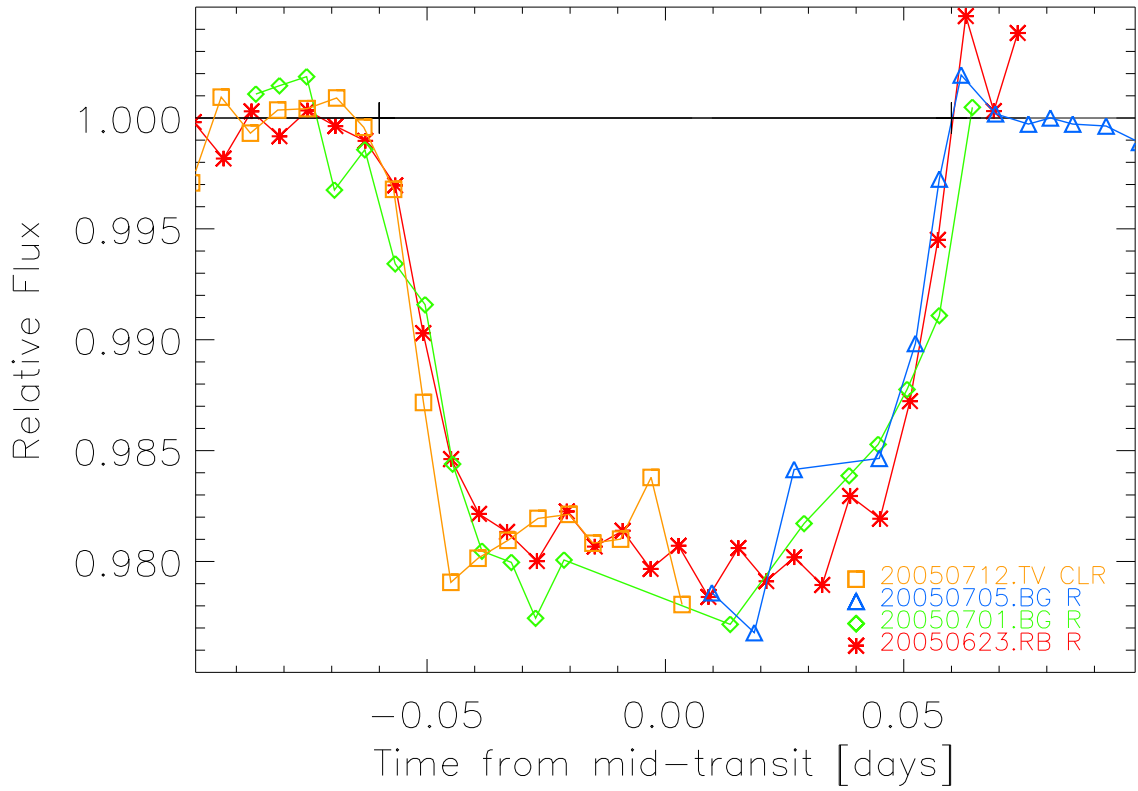}
\caption{Time series photometry of XO-1 from 2005, with dates, observers, and
filters indicated. The observations have been averaged in 0.006-day bins.
The figure is in color in the electronic edition.
\label{fig:etlc}}
\end{figure}

In June and July 2005, we observed XO-1 from three observatories in
North America and one in Europe. Telescopes of $\ga$0.2-m diameter have
adequate sensitivity for these purposes, and a network of them is well
suited to observe candidates with known positions and ephemerides.  A few
days after XO-1 was selected from the thousands of stars XO monitors,
the XO E.T. obtained light curves from independent telescopes sufficient
to discern a steep ingress and egress flanking a relatively wide flat
bottom (Figure \ref{fig:etlc}). With angular resolution $\sim$20 times
finer than the survey cameras', the E.T. observations reduced the
likelihood that XO-1 was a dilute triple system, i.e. one star with a
fainter, eclipsing binary nearby, which can mimic a transiting planet
(Charbonneau et al. 2004).

\begin{figure}
\plotone{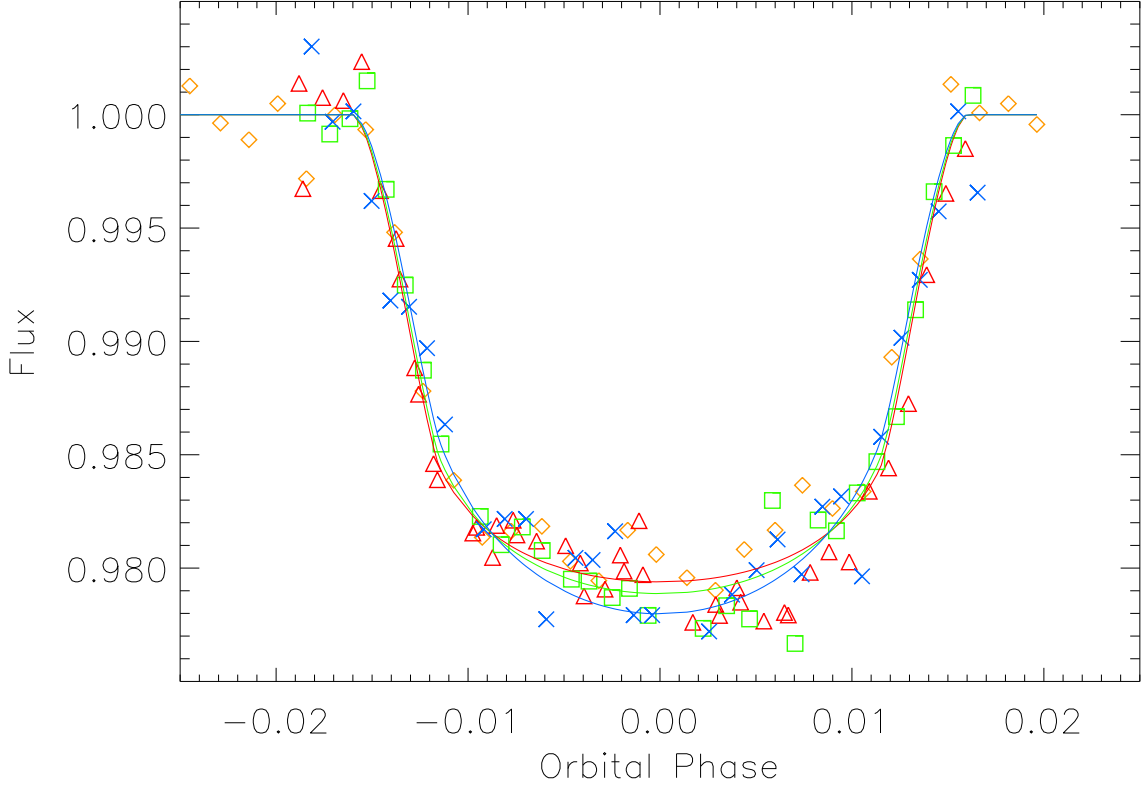}
\caption{Time series photometry of XO-1 from the Mar 14, 2006 transit
plotted on the same scale as Figure \ref{fig:etlc}.
We fit the R-band data [diamonds (0.3-m telescope; 0.006 day bins)
and triangles (1.8-m telescope; 0.002-day bins)]
to determine the system parameters for XO-1.
Although the shape of the R-band light curve provides the better
constraint on the inclination,
the V-band (squares) and B-band (crosses) data (both 1.8-m telescope)
appear to be slightly deeper and narrower than the R-band, as expected
for a near-central ($i \approx 90$\arcdeg) transit of a limb-darkened star.
The solid lines are models appropriate for each filter (color coded
in the electronic edition as B=blue, V=green, R=red) with
$m_* = 1.00$ \Msun,
$r_* = 1.00$ \Rsun,
$log~g = 4.53$,
$T_{eff} = 5750$ K, 
$r_p = 1.30$ \Rjup,
$i = 87.7$\arcdeg.
\label{fig:bestlc}}
\end{figure}

On the basis of the E.T. light curves of 2005, we acquired spectra of
XO-1 as described in the next section. After the first few spectra
showed XO-1b has substellar mass, we obtained additional photometry
(Figure \ref{fig:bestlc}) with the PRISM reimager (Janes et al. 2005)
in B, V, and R filters on the 1.8-m Perkins telescope at Lowell
observatory, and with the
E.T. telescopes. The photometry from the 1.8-m telescope was acquired
in a repeated exposure sequence of nine 5-s R images, three 10-s V images,
and one 30-s B image; the R and V data were averaged into 0.002-day bins.
The E.T. R data were averaged into 0.006-day bins.  Binning permitted
outlier rejection and empirical estimation of the noise by the scatter
of the individual observations, which is helpful in cases such as this
in which scintillation can be a significant and unpredictable component
to noise.

We estimate all-sky photometric B, V, ${\rm R_C}$, and ${\rm I_C}$ magnitudes
for XO-1 and
several nearby reference stars 
(Figure \ref{fig:allsky} and Table \ref{table:allsky})
calibrated using two Landolt areas (Landolt 1992).
Using a 0.35-meter telescope on photometric nights 2006 February 25 and
2006 March 14, we measured the fluxes of 28 and 22 Landolt stars, respectively
at an air mass similar to that for XO-1 
and established the zero points of the instrumental magnitudes and
transformation equations for the color
corrections for each filter and the CCD. 
The ${\rm R_C}$ magnitudes for XO-1 and 8 reference stars differed
on the two dates by a standard deviation of 0.005 mag.
The B, V, ${\rm R_C}$, and ${\rm I_C}$ absolute photometric accuracies
are 0.04, 0.03, 0.02, and 0.03 mag r.m.s.,
including both the formal error and an estimated systematic error. 
The Tycho magnitudes for XO-1 listed in 
Table \ref{table:star} transform (via Table 2 of Bessel 2000)
to Johnson $V = 11.25$, i.e. 0.06 mag (2-$\sigma$) fainter than our estimate.

\subsection{Spectroscopy}

In order to measure the mass ratio of the system and to determine the
characteristics of the host star, we obtained spectra of XO-1 with
two-dimensional cross-dispersed echelle spectrographs (Tull et al. 1995; 
Tull 1998) at the coude focus of the
2.7-m Harlan J. Smith (HJS) telescope and via a fiber optic 
on the 11-m Hobby-Eberly Telescope (HET).
Both telescopes are located at McDonald Observatory.
The HJS spectra were obtained in a traditionally-scheduled
manner, and the HET spectra were queued.
We used an iodine gas cell on the HET, but not on the HJS telescope.
We obtained one or two high-resolution ($\lambda/\Delta\lambda \approx 60000$)
spectra on each of ten nights. We extracted
the two-dimensional echelle spectra using procedures described in Hinkle et al.
(2000).

\section{Analysis}

\subsection{Ephemeris}

We refined our estimate of XO-1b's orbital period $P$ from modeling
eight pairs of transits chosen as follows:
each of four transits observed by the E. T.
on 2005 June 23, 2005 July 01, 2005 July 05, and 2005 July 12
(Figure \ref{fig:etlc}) was paired with two observations of
the 2006 March 14 transit (Figure \ref{fig:bestlc}). 
We solved for the best-fitting period for each pair of transits. The mean
orbital period, $P$ is \vperiod\ days; the 1-$\sigma$ dispersion is \eperiod\
days. The heliocentric Julian date of minimum light (also mid-transit) is
\begin{equation}
t_{m.l.} = t_c + P \times E,
\label{eq:ephem}
\end{equation}
where E is an integer, and $t_c = $\vjd$\pm$\ejd\ (HJD).
Identical parameters are listed in Table \ref{table:planet} and
certainly will be refined in subsequent publications.
In estimating the parameters of Equation \ref{eq:ephem},
we used only the photometry of the eight pairs of transits mentioned previously.
The XO survey photometry gives similar parameters but with larger
uncertainties, as expected because the survey data (by design) have
poorer photometric precision and cadence than the E. T. data.

\subsection{Radial Velocity Measurements}

\begin{figure}
\plotone{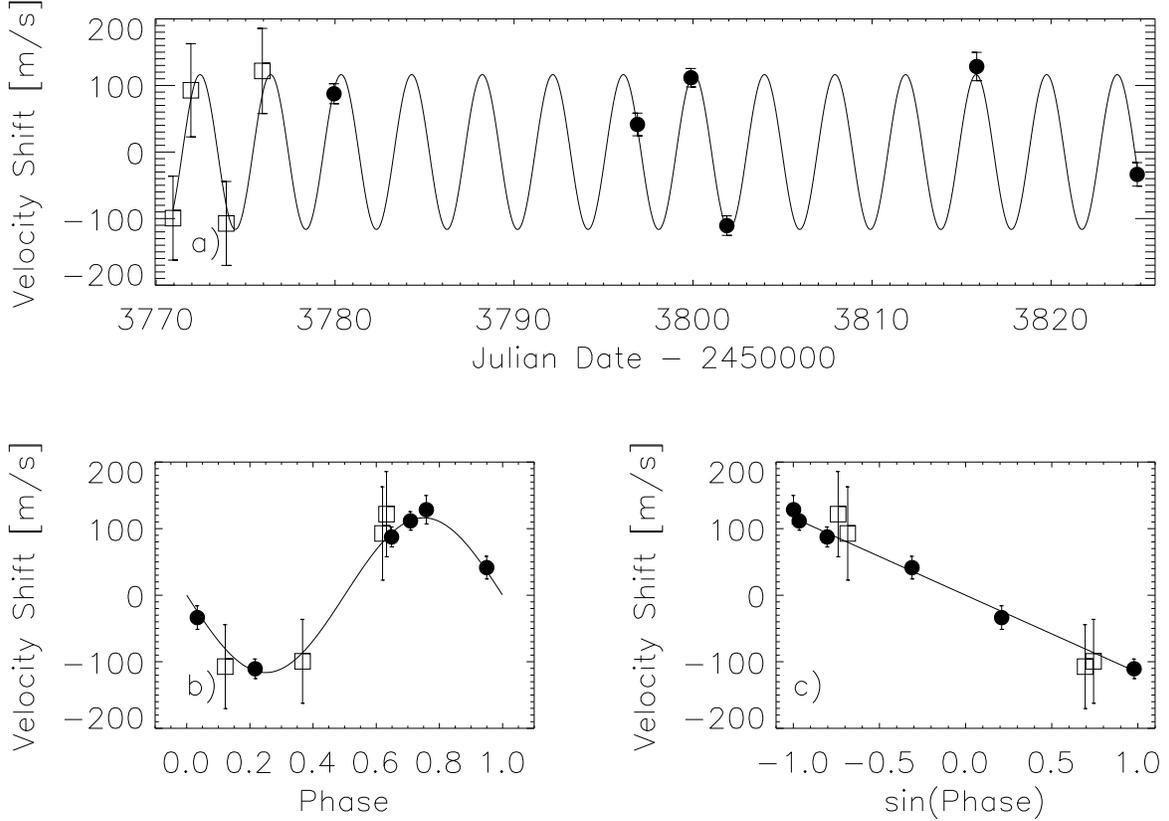}
\caption{a) The radial velocity of XO-1 oscillates sinusoidally with a
semi-amplitude K = \vrvK$\pm$\ervK\ \mps, implying XO-1b's mass is
\vMp$\pm$\eMp\ \Mjup. b) The
period and phase of the radial velocities were fixed at values determined
by the transits. The mean stellar radial velocity with respect to the
solar system's barycenter has been subtracted.  In order to determine
K, we used only the HET data calibrated with an iodine absorption cell
(filled circles).  The data obtained with the 2.7-m HJS telescope and
calibrated with ThAr emission lines (open squares) are for illustration
only. We subtracted mean radial velocities separately from the HET and HJS data.
Because the data (filled circles and open squares) are from independent
observers, telescopes, calibration sources, and analyses, we are confident
that the radial velocity oscillation is genuine. c) In this representation of
the data, a circular orbit yields a straight line of slope $-$K.
\label{fig:rviodine}}
\end{figure}

We measured XO-1's radial velocities with respect to the topocentric
frame using iodine absorption lines superposed on the HET spectra of XO-1.
We modeled the extracted spectra using high-resolution spectra
($\lambda/\Delta\lambda \approx 10^6$) of the Sun and the Earth's
atmosphere (Wallace, Hinkle, \& Livingston 1998)
and the HET iodine gas cell (Cochran 2000). Using an
IDL\footnote{IDL is a registered trademark of Research Systems, Inc.}
implementation of Nelder and Mead's (1965) 
downhill simplex $\chi^2$ minimization algorithm, ``Amoeba,'' we adjusted
parameters of our model to fit the observations.
The model includes convolution of our input spectra with
a best-fitting Voigt profile to approximate the (slightly non-Gaussian)
line-spread-function of the instrument.  The free parameters of our model
are a continuum normalization factor, the radial velocity of the star,
the radial velocity of the
iodine lines (which represent instrumental deviations
from their expected zero velocity with respect to the observatory),
and an exponent (optical depth scale factor) that ``morphs'' the depths of
the lines, as an arbitrary method of adjusting the solar
spectrum to that of XO-1.  Due to the iodine absorption, we could not
estimate the continuum level by interpolating between local maxima
in the spectrum, so instead we solved for the continuum iteratively,
as required to improve the fit between our model and the observations.
In the manner described above, for each $\sim$15 \AA\ section of each
individual spectrum within the region of the recorded spectrum with
significant iodine absorption, $5100 - 5700$ \AA,
we estimated the radial velocity of the star.  From the approximately
normal distribution of the resulting radial velocity estimates for each
epoch, we
calculated the mean radial velocity and its uncertainty.
We transformed our measured radial velocities to the barycentric frame
of the solar system and phased them to the
ephemeris known from the transits (Figure \ref{fig:rviodine} and
Table \ref{table:rv}).

The 1-$\sigma$ internal errors of the radial velocities from the iodine
spectra are $\sim$15 \mps\ per epoch (i.e. per average of two 15-minute
exposures) with the 11-m HET and $\sim$65 \mps\ per 15- or 20-minute exposure
with the 2.7-m HJS telescope.  We used only the iodine-calibrated
radial velocities to determine the amplitude of the radial velocity
variation, assumed to be sinusoidal and phased with the transits.
An eccentricity approximately equal to zero is expected theoretically
(Bodenheimer et al. 2001) and is consistent with the radial velocities
(Figure \ref{fig:rviodine}c).
The maximum likelihood radial velocity semi-amplitude
K = \vrvK$\pm$\ervK\ \mps.

\subsection{Spectroscopically-Derived Stellar Properties and Planetary Mass}

An initial estimate of the spectral type of XO-1 was obtained using
the line depth ratio method developed by Gray \& Johanson (1991).
The spectral type is estimated by measuring the depth of the \ion{V}{1}
6251.83 \AA\ line divided by the depth of the adjacent \ion{Fe}{1}
6252.57 \AA\ line.  The 4 spectra from the HJS telescope were summed
together and the individual line depths were measured by fitting a
Gaussian function to each line.  The resulting line ratio is 0.14,
which corresponds to a G1 star (Table 1 of Gray \& Johanson 1991).
Differences in metallicity can affect these results (Gray 1994), so
we estimate an uncertainty of one spectral subclass for the spectral
type of X0-1.  These published line depth ratios are for dwarf stars,
and Gray \& Brown (2001) show that there is a gravity dependence to the
spectral type classification that results from this ratio. We
estimate the luminosity class V by examining the gravity-sensitive
Na D doublet and the Mg b triplet.
Below we estimate the log$g = 4.53\pm0.065$\ \cmpss, whereas early G-type
luminosity class III giants have log$g \approx2.5$\ \cmpss, which produces
substantially narrower Na D profiles (Gray 1992).
Our final spectral type estimate for XO-1 is \sptype.

We used the software package SME (Valenti \& Piskunov 1996) to fit each
of the four spectra of XO-1 from the HJS telescope with synthetic spectra.
We used
the methodology of Valenti \& Fischer (2005), including their minor
corrections to match the Sun and remove abundance trends with temperature
(negligible in this case). Because of gaps between echelle orders,
the HJS spectra are missing the wavelength intervals 6000--6123
\AA\ and 6113--6118 \AA, which were included in the Valenti \& Fischer
analysis. These wavelength intervals are also missing from our extracted
HET spectra because the relevant echelle orders span the gap between
the two detectors.

We averaged our SME results for the four HJS spectra, obtaining
the parameter values in Table \ref{table:sme}. 
Each value in the last column of the table, labeled ``Precision'' because
systematic uncertainties are not included,
is the standard deviation of the four measurements divided
by $\sqrt{3}$ to yield the formal uncertainty in the mean.
The median value of
each derived parameter (not given) differs from the mean by less than the
uncertainty in the mean. The final row in the table gives [Si/Fe], which
Valenti \& Fischer (2005) used as a proxy for alpha-element enrichment,
when interpolating isochrones.

\begin{figure}
\epsscale{0.8}
\plotone{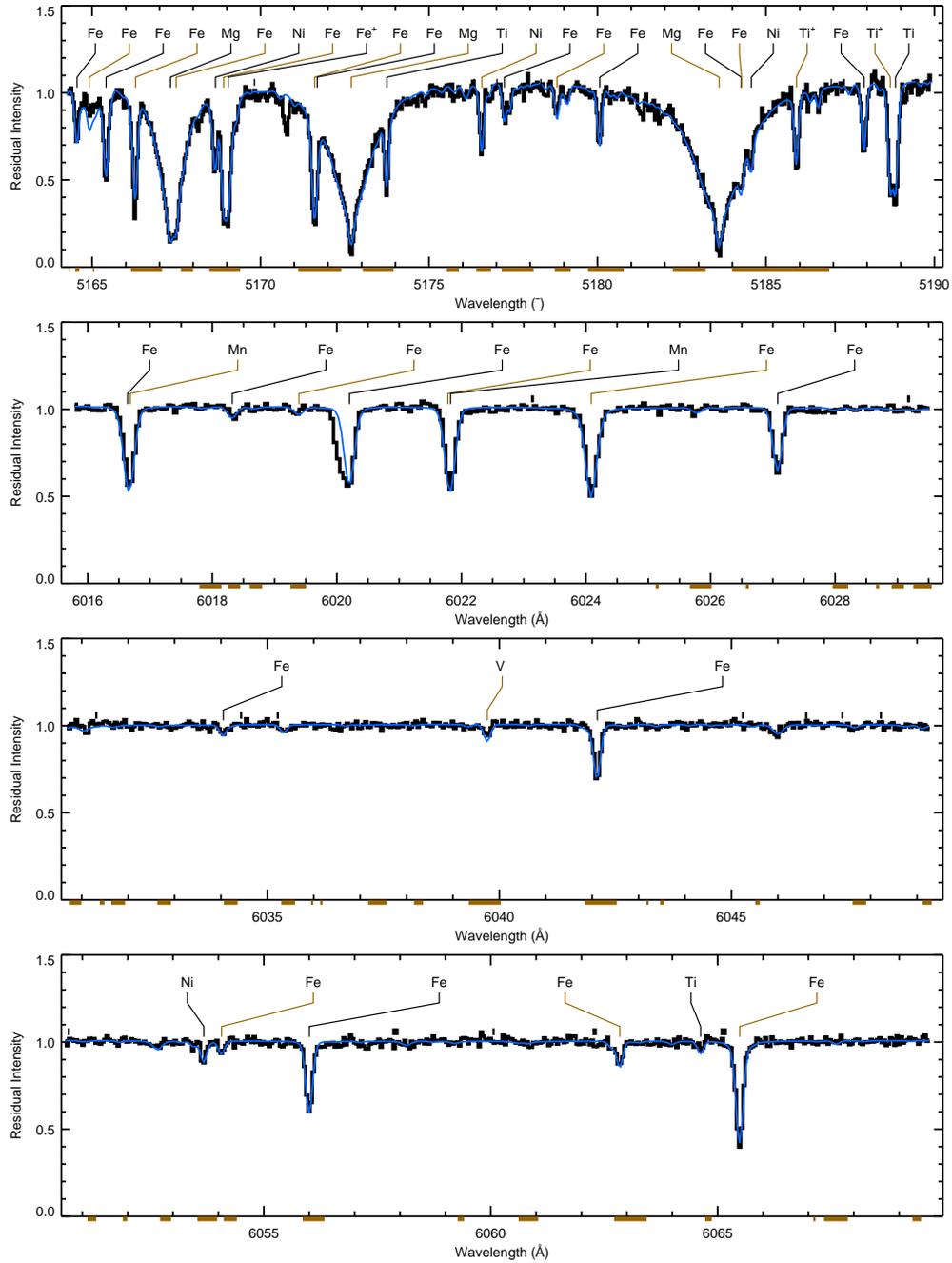}
\caption{
The mean spectrum of XO-1 as observed (black
histogram) and modeled with SME (curve, colored blue in the electronic edition)
in the region of the Mg b triplet.
Labels note the elements responsible for the indicated spectral lines.
Intermittent line segments (tan) beneath the horizontal
axis indicate wavelength intervals used to constrain the
spectroscopic parameters. Very short and intermittent line segments (black)
immediately above
the spectrum indicate wavelength intervals used to constrain the
continuum fit.
\label{fig:mcd}}
\end{figure}

Figure \ref{fig:mcd} shows XO-1's spectrum in the region of the Mg b
triplet, which is the dominant spectroscopic constraint on gravity. These
three Mg lines also have a significant impact on the global [M/H]
parameter, which is used to scale solar abundances for all elements
other than Na, Si, Ti, Fe, and Ni.  The unidentified absorption lines
at 5171 \AA\ and in the wing of the 6020 \AA\ line are common in stars
of this type (Valenti \& Fisher 2005) so presumably are unrelated to
the presence of a hot Jupiter.

Following the methodology of Fischer \& Valenti (2005), we interpolated
Yonsei-Yale (Y$^2$) isochrones (Demarque et al. 2004) to determine
probability distribution functions for the mass, radius, gravity, and age
of XO-1. The trigonometric parallax of XO-1 is unknown, so we initially
assumed distances of 180, 200, and 220 pc with an adopted uncertainty of
10 pc in each case (which affects the width of the resulting distribution
functions). We measured a V magnitude of 
$11.19\pm0.03$ for XO-1 (Table \ref{table:star})
so XO-1 would be at 190 pc,
if it has the same intrinsic brightness as the Sun.

\begin{figure}
\plotone{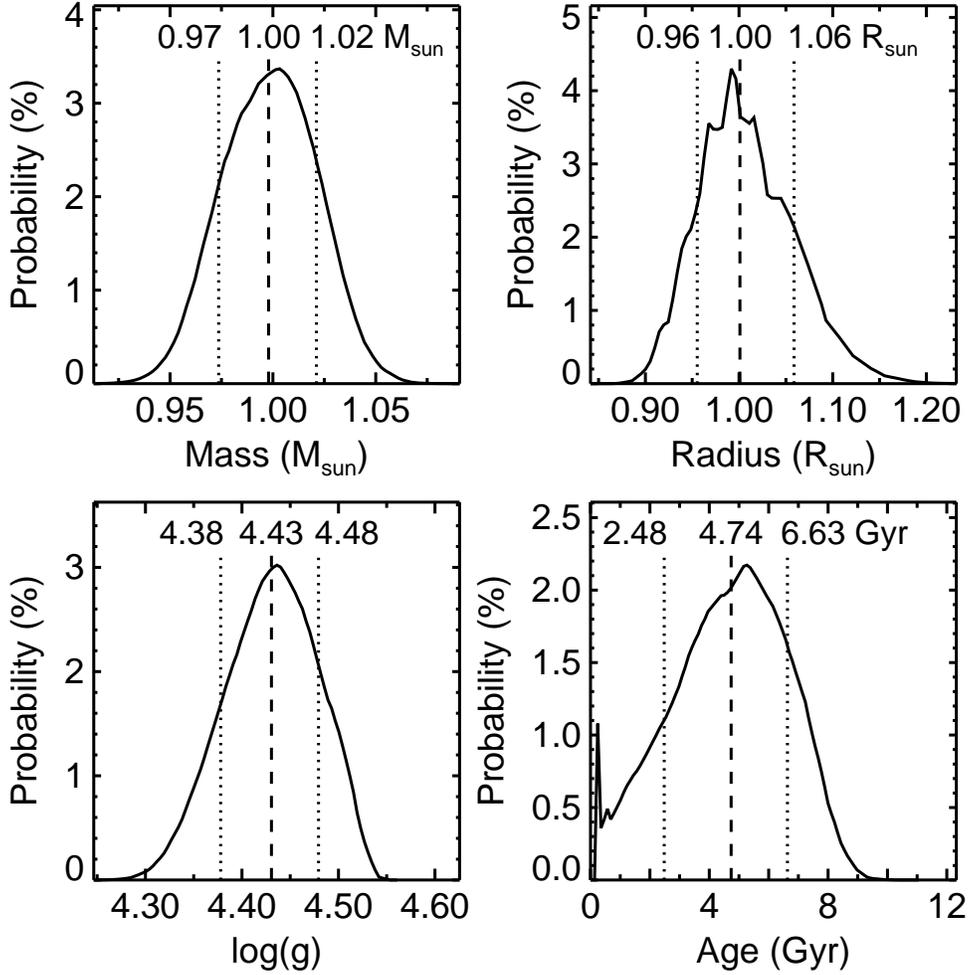}
\caption{Distributions for four stellar parameters derived from the
SME analysis (see text) for a distance of 200 pc. The values of
the mean and limits containing $\pm34$\%\ of the distribution from the mean
are annotated on the figures and listed in
Table \protect{\ref{table:stellarparam}} along with
corresponding values for distances of 180 pc and 220 pc.
\label{fig:stellarparam}}
\end{figure}

We used our spectroscopic effective temperature, spectroscopic gravity,
and an assumed distance to derive a bolometric correction by interpolating
the ``high temperature'' table from VandenBerg \& Clem (2003).
We combined the bolometric
correction with the observed V-band magnitude to determine stellar
luminosity. Then we used the stellar luminosity and our spectroscopic
effective temperature, iron abundance, and alpha-element enrichment to
interpolate the Y$^2$ isochrones to produce the probability distribution functions
in Figure \ref{fig:stellarparam}.

As expected, the most probable mass, radius, gravity, and age are
essentially solar for an assumed distance of 200 pc. If we assume that
XO-1 is at 180 pc instead, then the most probable age drops to 2 Gyr,
which is unlikely given how slowly XO-1 rotates (V~sin$i < 3$\ \kps).
If we assume XO-1 is at a distance of 220 pc, then the gravity from
stellar structure models deviates significantly from the gravity we
derived spectroscopically. Thus, we assume in subsequent analysis that
XO-1 is at a distance of 200 pc, but clearly the trigonometric parallax
needs to be measured.

\begin{figure}
\plotone{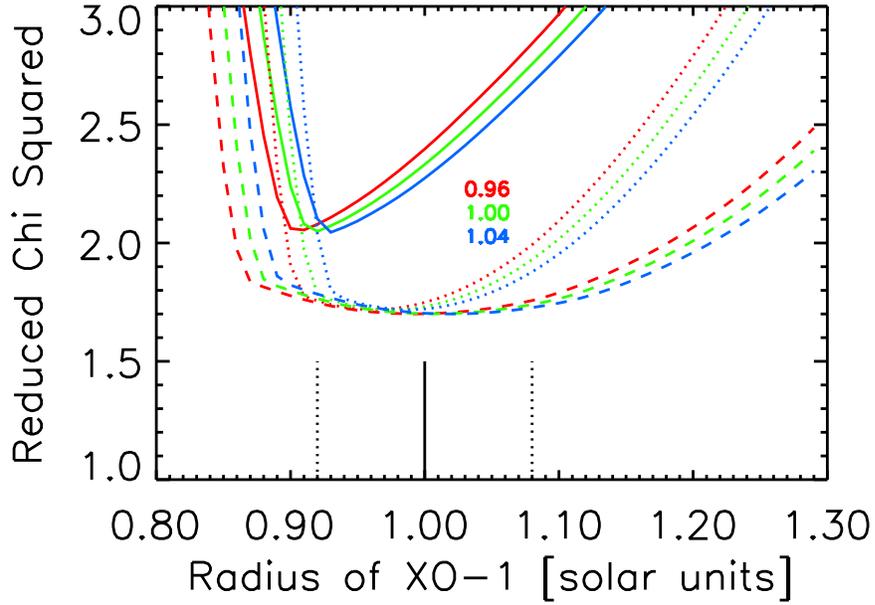}
\caption{The reduced $\chi^2$ of light curve models rises away from $\sim$1
solar radius, more steeply at smaller radii than larger radii
because the cosine of the inclination cannot
exceed unity (cf. Equation \ref{eq:incl}). Three independent observations of 
two transits are shown as solid, dotted, and dashed lines. The nearly coincident
lines correspond to $m_* =$ 0.96, 1.0, and 1.04 \Msun\ (colored in the
electronic edition). Because the mean stellar density is the critical parameter
(Equation \ref{eq:incl}), at a specific $\chi^2$, the more massive models
require larger stellar radii. 
The vertical solid line at $r_* = 1$\Rsun\ is the maximum likelihood value from
the SME analysis (Table \ref{table:stellarparam}); the vertical dotted lines
are the corresponding 1-$\sigma$ limits.
\label{fig:chisq}}
\end{figure}

Assuming a distance of $200\pm10$ pc yields a 68\% confidence limit
of $1.00\pm0.025$ for M/\Msun, $1.00\pm0.05$ for R/\Rsun, and
2.5-6.6 Gyr for the age (Figure \ref{fig:stellarparam}).
The stellar mass combined with our measured semi-amplitude
K = \vrvK$\pm$\ervK\ \mps\ yields 
a mass of \vMp$\pm$\eMp\ \Mjup\ for the planet XO-1b.
Within the broader range of distances of
$200\pm20$ pc (Table \ref{table:stellarparam}), 
the light curves and the spectroscopy produce consistent
estimates of the stellar radius (Figure \ref{fig:chisq}).
Based upon XO-1's proper motion, radial velocity, and distance,
its velocity vector with respect to the local standard of rest is 
(u,v,w) = (8, 9, 23) \kps, which is consistent with
the kinematics of a G-dwarf with an age of a few Gyr (Wyse \& Gilmore 1995).

\subsection{Light Curve Modeling and Planetary Radius}

We modeled the observed transit light curves using the algorithms of
Mandel \& Agol (2002) and adopt their notation. For the star, we
interpolate quadratic limb darkening coefficients from Claret (2000) for the
filter bandpasses:
B ($\gamma_1 = 0.639$; $\gamma_2 = 0.179$),
V ($\gamma_1 = 0.437$; $\gamma_2 = 0.295$),
or
R ($\gamma_1 = 0.339$; $\gamma_2 = 0.324$),
based upon our spectroscopic gravity and effective temperature and assuming
solar metallicity, which is consistent with
our spectroscopic determination (Table \ref{table:sme}).
In addition to the two limb darkening coefficients, parameters of a model
light curve include the mass $M_*$ of the star, the size ratio $p =
r_p / r_*$ of the radius of the planet $r_p$ to the radius $r_*$ of
the star, the inclination $i$ and the angular frequency $\omega$ of
the orbit, and the heliocentric epoch of the transit's center, $t_c$.
Note that the shape depends on the size ratio $p = r_p / r_*$ and not
the two radii separately.

Of the seven input parameters ($\gamma_1$, $\gamma_2$, $M_*$, $p$, $i$,
$\omega$, $t_c$) the most precisely measured are $\omega$ and $t_c$.
As previously noted, we estimated $\gamma_1$ and $\gamma_2$ from the
spectroscopically-determined log$g$ and $T_{eff}$.  Our derivation of the
other parameters is similar to that of Brown et al. (2001).  We begin
by adopting the stellar mass $M_*$ and its uncertainty from the previous
section. Combining the
mass $M_*$ and the angular frequency $\omega$ with an assumption of a
circular orbit gives the velocity of the planet
passing in front of the star. Combining that velocity with the observed
duration of the transit (from 2nd contact to 3rd contact) gives a chord
length for the transit across the star.  The half chord length $L$ must be
less than or equal to the radius of the star, i.e. $L \le r_*$.
The stellar radius $r_*$ also has an upper bound from the light curve: for
 a transit of a given $L$, the larger $r_*$ becomes, the
longer will be the ingress and the egress, as the transit becomes more
grazing.  Thus, the duration from 1st to 2nd contacts (or 3rd to 4th)
creates an upper bound, $r_* \la 1.3$\Rsun, whereas the duration from 2nd
to 3rd contacts creates a lower bound, $r_* \ga 0.9$\Rsun.

For rhetorical purposes we have discussed the transits as if the contacts
are distinct features in the light curves, although with the precision of our
photometry, they are not.
Instead, the model light curves become too narrow compared to the
observations if $r_* \la 0.9$\Rsun, and the models become too V-shaped to
match the ``steep-edged'' U-shape of the observations if $r_* \ga 1.3$\Rsun.
In between these extremes, $r_*$ and $i$ are anti-correlated: increases
in $r_*$ can be accommodated by decreases in $i$, and vice versa:

\begin{eqnarray}
i & = cos^{-1}\sqrt{ \theta_\odot^2 \hat{P}^{-4/3} \hat{\rho}^{-2/3} - \pi^2 q^2}\nonumber\\
& \approx {\pi\over{2}} - \sqrt{ \theta_\odot^2 \hat{P}^{-4/3} \hat{\rho}^{-2/3} - \pi^2 q^2}
\label{eq:incl}
\end{eqnarray}
where $\theta_\odot = 1 $\Rsun$/1 {\rm A.U.} = 4.65\times 10^{-3}$ rad,
$\hat{P}$ is the planet's orbital period in years,
$\hat{\rho}$ is the dimensionless mean density of the star in solar
units i.e. $\hat{\rho} = (m_*/$\Msun$)(r_*/$\Rsun$)^{-3}$, and
we determine the scalar $q$ so as to match the relation between 
each best-fitting inclination $i$ for each assumed value for $\hat\rho$.
Geometrically, $q$ is the dimensionless duty cycle of the transit,
i.e. the duration of the transit divided by the period,
where the transit's duration is defined as the interval between the
two moments at which the planet's center coincides with the limb of the star.
For XO-1, $q = $\valq$\pm$\errq,
$r_* = $\vRs$\pm$\eRs \Rsun,
$m_* = $\vMs$\pm$\eMs\ \Msun, so
$\hat{\rho} = 1.0\pm0.24$ and thus $i = $\vincl\arcdeg$\pm$\eincl\arcdeg.

The radius of the planet $r_p$ is approximately proportional to $r_*$,
because the depth of the transit is approximately equal to $p^2$. At a more
precise level of approximation,  $r_p \propto r_*^\kappa$, where $\kappa$
accounts for limb darkening, which necessitates an increasingly larger $r_p$
as $r_*$ increases (Equation \ref{eq:r_p}). 
For various
presumed values for $r_*$ and $m_*$, we minimized $\chi^2$ of the R-band
data by varying $t_c$, $r_p$, and $i$.  The region of high likelihood
for XO-1 is described by the functional relation,
\begin{equation}
r_p \approx (1.30\pm0.03) (r_*/r_\odot)^{\kappa} {\rm R_J},
\label{eq:r_p}
\end{equation}
where $\kappa \approx 1 + 0.5 \gamma_1 L / r_*$. As previously defined,
$\gamma_1$ is the linear limb darkening coefficient, and $L$ is the half chord
length of the transit's path across the star. 
For XO-1 observed in R band, $\kappa = 1.17$.
Substituting the best-fit value and uncertainty for
the radius of the star into Equation \ref{eq:r_p}, we
conclude that the planet's radius\footnote{Authors unwittingly create
ambiguity if they do not define \Rjup\ explicitly, because various
averages of Jupiter's polar radius $R_p$
and its equatorial radius $R_e$ could be selected for \Rjup.
Because $R_p \approx 0.94 R_e$ (Mallama et al. 2000), 
such ambiguity can be significant compared to observational uncertainties;
e.g. Knutson et al. (2006) report the radius
of HD 209458 with a fractional uncertainty of 1.9\%.
From reported masses and mean densities,
we determined that many authors tacitly define \Rjup\ $= R_e = 71492$ km;
hence we have also, for the sake of consistency.}
as determined from the R-band light curves is \vRp$\pm$\eRp\ \Rjup.

Limb darkening is stronger in the blue than the red, so a transit with
inclination $i = 90$\arcdeg\ should be deeper in the blue than the red.
The difference in transit depths in B band compared to R band would be
2 mmag for XO-1b if $i = 90$\arcdeg.  At the other extreme, a grazing
transit, the transit should be shallower in the blue than the red.
The transit depth color difference's constraint on inclination requires
more accurate photometry than that presented here (cf. Jha et al. 2000).

\section{Discussion}

\subsection{Transit Timing}

Other planets in the XO-1 system could be revealed by their transits or
by their perturbations of the radial velocity of XO-1 or on the
time of arrival of the transits by XO-1b. We have not observed sufficient
numbers of transits by XO-1b with adequate quality to justify an analysis
of time-of-arrival perturbations. To motivate such observations, we note
their scientific value and present an analysis of the dependence of timing
precision on photometric precision.

Transit timing observations would be
especially sensitive to planets in orbital resonance(s) with XO-1b
(Steffen \& Agol 2005; Holman \& Murray 2005). 
In principle the error in a measurement of the time of arrival of a transit
is proportional to the photometric error per unit time. We estimate that
for an idealized transit light curve, with an isosceles
trapezoid shape, i.e. with negligible limb darkening, the error
in the time of arrival (TOA) of each transit is
\begin{equation}
\sigma^2(TOA) \approx ({{\sigma_m~w}\over{d \sqrt{N}}})^2 + ({{\epsilon~w}\over{2~d}})^2
\label{eq:toa}
\end{equation}
where $\sigma_m/d$ is the dimensionless ratio of the standard deviation of
the photometry to the depth of the transit, $N$ is number of observations
obtained during ingress and/or egress, $w$ is the time interval between
first and second contacts,
and $\epsilon$ is a characteristic error due to (uncalibrated) secular
variations in gain or atmospheric transparency over a time scale equal
to $w$.  Implicit in Equation \ref{eq:toa} is that 1) the light curve
shape is known, 2) there are sufficient observations to set the level
of the light curve's out-of-transit portion precisely, 3) the noise has
a normal distribution, and 4) N $\gg 1$. For example, if $\epsilon =
0$ and one can obtain $\sigma_m = $ 2 mmag precision per 1-minute observation
of a transit with depth $d = 20$ mmag and duration of ingress $w = 20$
minutes, then for an observation of ingress, $N = 20$ and the time can be
measured with a standard deviation $\sigma(TOA) = 27$ seconds.
If ingress and egress are both observed, then $N = 40$ and $\sigma(TOA) = 19$
seconds for the transit as a whole. For $\epsilon = \pm$1 mmag
the measured time of arrival of the transit is shifted by $\pm$30 seconds.
If there
are additional uncalibrated systematics in an observed light curve,
then Equation \ref{eq:toa} will underestimate the uncertainty of the
time of arrival. In particular, if only ingress or egress is observed,
then secular trends are more difficult to calibrate and the residual error in
calibration, $\epsilon$, will be larger than if many observations are
obtained immediately before and after each transit.

Because scintillation is independent of the brightness of the star,
the photometric precision, and hence the timing precision of transits
observed from the Earth's surface, is not strongly dependent of the
brightness of the target star, for stars as bright or brighter than XO-1.
If the star is sufficiently faint that the Poisson contribution to
noise is much larger than the scintillation, e.g. if the sensor or the
spectral filters limit the collecting efficiency, such as in B band, or
if spectroscopy is the goal, bright stars and/or large telescopes can
have a significant advantage.  The Poisson noise for XO-1 observed with
a 0.3-m telescope with an R-band filter is 0.15\% (i.e. $\sim$1.5 mmag)
in a 60-second exposure. Near sea level, typical scintillation for such a
telescope and exposure time is 0.9, 1.7, and 2.9 mmag at air masses of 1,
1.5, and 2, respectively (Young 1967; Dravins et al. 1998). 
For differential photometry, comparison stars comparable in brightness to
the target star generally will be nearer to a faint target than 
a bright target. The latter effect
can be important if the nearest suitable comparison star is
too distant in angle to observe with the target simultaneously on the same
sensor.

We searched for a second planet transiting XO-1 by removing the
observations within 0.06 days of minimum light of the known transit.
The sensitivity to transits decreases with increasing period, and
the a priori likelihood that a planet orbiting XO-1 will transit also
decreases with period.  Due to the measurement errors of the photometry
from the XO cameras and the radial velocities, respectively, we would not
have detected a second planet orbiting XO-1 if its radius $r_p \la 0.5$
\Rjup\ or if its mass $m_p$, inclination $i$, and angular frequency, are
such that $m_p~sin(i)~{(\omega_2/\omega_1)}^{1/3} \la 0.5$ \Mjup, where
$\omega_2/\omega_1$ is the ratio of the angular frequency of the second
planet to that of XO-1b. The complementary statement, that we would have
detected a second planet if either or both of those inequalities were not
true, cannot be definitive in general. In particular, additional planets
could orbit XO-1 with periods much longer than that of XO-1b. Monitoring
XO-1's radial velocity could reveal additional planets and enable predictions of
the approximate times that they might transit.

\subsection{A Triple Star?\label{sec:triple}}

Mandushev et al. (2005) demonstrate that a triple star system (gravitationally
bound or an asterism) can exhibit
a light curve and a radial velocity curve similar to that of a planet
or brown dwarf transiting a single star.
The rotational broadening of Mandushev et al's stars, (V~sin$i$ = 34\ 
\kps\ for the F0 primary; V~sin$i$ = 15\ \kps\ for the G0 secondary)
produced an oscillating asymmetry in the spectral lines that could be
misinterpreted as periodicity in the radial velocity of a single star
induced by a dark companion of substellar mass.  We performed a simpler
form of Mandushev et al's analysis for XO-1 which we summarize here.
Our analysis is simpler because XO-1 has narrow spectral lines (V~sin$i < 3$\ \kps),
so perturbations by a stellar companion would be observed as shifts of
narrow lines, rather than distortions of broad lines.

Mandushev et al's light curve consists of three parts
equal in duration: ingress, flat bottom, egress.  Although a transiting
planet can produce such a light curve, the a priori probability of
the duration of ingress (or egress) being as large or larger than the
duration of the transit's middle section, between contacts 2 and 3, is
approximately $1 - \sqrt{1 - 2 r_p / r_*}$, which is always greater than
or equal to $p = r_p / r_*$, and equals $\sim$10\% for a typical Jovian
planet transiting a solar-type star.  In practice, we expect such grazing
transits to be detected with lower than a priori chance, because their
reduced depth and duration both reduce detectability. On the other hand,
if $p \ge 0.375$, the probability of ingress (or egress) being at least
as long as the middle section of the light curve is $\ge 0.5$;
that is, eclipsing binary stars typically exhibit such light curves,
whereas transiting planets typically do not. Such probabilistic arguments
cannot prove a given system is not a transiting planet, but we have used
them to focus our efforts on ``steep-edged'' U-shaped light curves instead
of those with ingress and egress comparable in length to the flat bottom.

For XO-1's light curve, the total duration of the flat bottom is $\sim$5 times the
duration of ingress or egress, which implies that the ratio of diameters
of the eclipsing bodies must be $\ga 5$.  For a G1 V primary ($r_*=1.0$ \Rsun),
an eclipsing stellar companion must have a radius $r_* \la 0.2$ \Rsun\ to match
the duration of ingress relative to the flat bottom, which implies a spectral
type no earlier than M5 V. Diluting the transit depth with a second
G1 V star in the system would mimic the observed depth (2\%) without affecting
the durations.\footnote{In this hypothetical example, we could replace the
M5V star with a planet twice the nominal
mass of XO-1b and twice the latter's projected surface area, and all
the observational constraints would be equally well satisfied. However,
the density of the planet would be $1/\sqrt{2}$ times the nominal
(single-star hypothesis) value.  Such a contrived example seems less
plausible than the nominal system of a single star and planet.}
However, the radial velocity
shifts of a G1 V star orbited by a M$\approx 0.1$\Msun\ companion
would be $\sim$10\ \kps. For a star with narrow lines like XO-1, velocity
shifts of this amplitude would be apparent in our spectra, in contrast to
the situation of stars with much broader spectral lines.

Modeling XO-1 as a triple star with a second star fainter than the G1V primary
encounters difficulty with the radius
of the third star (that eclipses the second star). 
As the second star contributes a lesser fraction of
the total light of the system, we are forced to make the third star
an ever larger fraction of the size of the star it eclipses, in order to
match the 2\% depth.  However, from the light curve we know that the
ratio of diameters of the eclipsing bodies must be $\ga 5$.  

Modeling XO-1 as a triple star cannot include a secondary star brighter than
the observed narrow-lined \sptype\ star, unless 
the secondary star's spectrum is featureless, perhaps due to great
rotational broadening. However, the broadband colors of XO-1 are consistent
with a slightly reddened\footnote{In the direction of XO-1,
the interstellar extinction
${\rm A_V = 0.04}$ mag as inferred from infrared observations with
a 6\arcmin\ (FWHM) beam (Schlegel, Finkbeiner, \& Davis 1998).
We neglect its small ($\la$2\%) effect on the distance and the radius of XO-1.}
$T_{eff} = 5750$ K star (Table \ref{table:star}).

We conclude
the XO-1 transit cannot be caused by a stellar binary diluted by a third star
that is either physically bound or along the line of sight, due to
the mutual constraints of eclipse depth, eclipse shape,
and a single-lined \sptype\ spectrum with V~sin $i < 3$\ \kps. 

XO-1b must be a planet, but the XO-1 
system could include a wide-separation stellar companion that is
either much fainter than or very similar in spectra to the \sptype\ star.
For example, HD 189733 has a faint stellar companion (Bakos et al. 2006b).
For XO-1, the possibility of a stellar companion could be tested with
infrared adaptive optics on the largest available telescopes.

\subsection{Comparison to other Transiting Planets\label{sec:cf}}

Figure \ref{fig:rm} compares the radius and mass of XO-1b with those of the
nine other transiting extrasolar planets as tabulated by
Charbonneau et al. (2006). 
The mean density of XO-1b is 0.51$\pm$0.13 g cm$^{-3}$.
The radius of the star XO-1
is not known to sufficient accuracy to assert that its planet XO-1b is truly
anomalously large. If the G1 V star XO-1 has the same radius as the Sun,
then the planet XO-1b is approximately the size of HD 209458b. 
The radius of each of these two large planets,
$\sim$1.3 times the radius of Jupiter, is $\sim$20\% larger
than theoretical models predict (Table 1 of Bodenheimer et al. 2003).
To keep the planet's size large, an additional heat (or power) source
is hypothesized deep inside the planet.
Knutson et al. (2006 and references therein) summarize
three proposed models for that heat source: 1) strong winds on the planet,
with the original power source being the incident light of the star,
2) tidal dissipation, presumably due to a second planet forcing the hot
Jupiter into an eccentric orbit, and 3) resonances between the spin precession
and the orbital precession, a so-called Cassini state.
At \vMp\ \Mjup, XO-1b is $\sim$2 times more massive than the
best-fit correlation of mass and period noted by Mazeh, Zucker, \& Pont (2005),
whereas HD 149026b is $\sim$2 times less massive than the same correlation.

\section{Summary}

\begin{figure}
\plotone{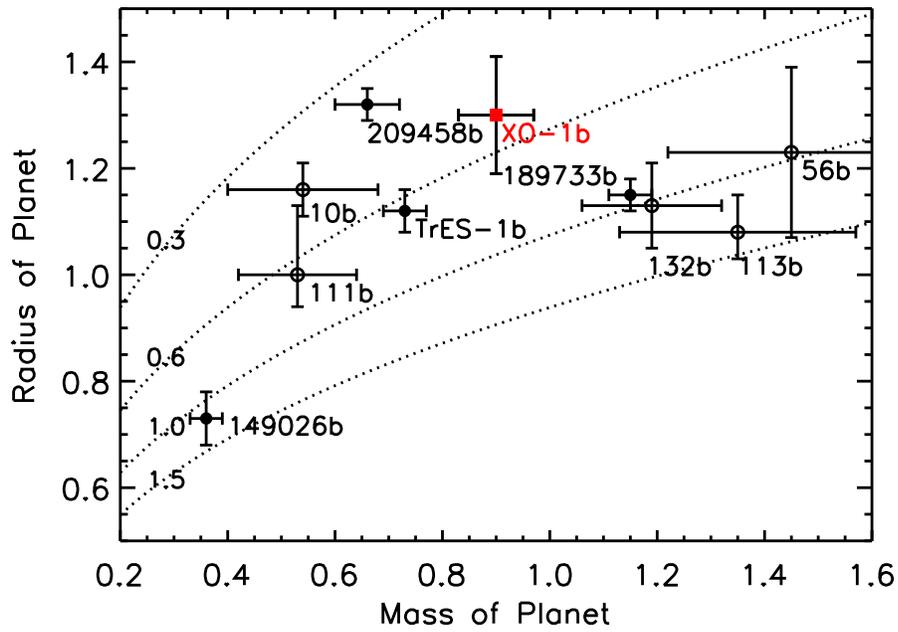}
\caption{To this plot of the radii and masses of transiting extrasolar planets
in Jovian units (cf. footnotes to Table \ref{table:planet}),
we have added XO-1b (red square) to
the latest data for the other nine compiled by Charbonneau et al. (2006).
Filled symbols are stars brighter than V=12. Open circles represent
OGLE planets (V$\sim$15).
Lines of constant mean density are labeled in g cm$^{-3}$.
\label{fig:rm}}
\end{figure}

The planet XO-1b is the first planet discovered by the XO project. 
It is similar to other transiting extrasolar planets in its
physical characteristics
(Section \ref{sec:cf}; Table \ref{table:planet}; Figure \ref{fig:rm}).
Of those nine others, HD 209458b has the most similar radius to the
nominal radius of XO-1b.
The host star XO-1 is very similar to the Sun in its physical characteristics
(Tables \ref{table:star}, \ref{table:sme}, and \ref{table:stellarparam}).
Equation \ref{eq:incl} expresses the inclination of a transiting
planet's orbit as a function of its period, the duty cycle of the transit,
and the mean density of the star.
For transits with imprecisely measured inclinations near 90\arcdeg,
the derived ratio of the radii of the planet and the star
is approximately proportional 
to the stellar radius raised to a power equal to one half the linear
limb darkening coefficient (Equation \ref{eq:r_p}).
We note the scientific potential of transit timing
and estimate limits to its accuracy. 
We discuss evidence that disproves the hypothesis that the periodic
modulations of the photometry and the radial velocity of XO-1 are due
to a triple star system.

\acknowledgments

The University of Hawaii staff have made the operation on Maui possible; we
thank especially Bill Giebink, Les Hieda, 
Jake Kamibayashi,
Jeff Kuhn, Haosheng Lin, Mike Maberry,
Daniel O'Gara, 
Joey Perreira, Kaila Rhoden, and the director of the IFA, Rolf-Peter Kudritzki.

We thank
Lisa Prato, Marcos Huerta, Danielle Best, and Josh Shiode for assistance observing.
We acknowledge helpful discussions with 
Gaspar Bakos,
Christopher Burke,
Fred Chromey, 
Ron Gilliland, 
Leslie Hebb,
James McCullough,
Margaret Meixner,
and
Kailash Sahu. 

This research has made use of a Beowulf cluster constructed by Frank
Summers; the SIMBAD database, operated at CDS, Strasbourg, France;
data products from the Two Micron All Sky Survey (2MASS) and
the Digitized Sky Survey (DSS);
source code for transit light-curves (Mandel \& Agol 2002);
and community access to the HET.
We thank the support staff of the McDonald Observatory.

XO is funded primarily by NASA Origins grant NAG5-13130, and in the past
has received
financial support from the Sloan Foundation, the Research Corporation,
the Director's Discretionary Fund of the STScI, and the US National
Science Foundation.

\clearpage
\begin{deluxetable}{ccccc}
\tabletypesize{\small}
\tablewidth{0pt}
\tablecaption{{\rm All-sky Photometric Magnitudes}}
\startdata
\hline
\hline
Star${\rm ^a}$	&B 	&V	&${\rm R_C}$	&${\rm I_C}$ 	\\
\hline
XO-1 	&11.85	&11.19	&10.806	&10.43 \\
   1 	&13.85	&12.85	&12.282	&11.76 \\
   2 	&16.31	&14.93	&14.139	&13.41 \\
   3 	&12.36	&11.54	&11.046	&10.56 \\
   4 	&14.31	&13.65	&13.313	&12.93 \\
   5 	& 9.99	& 9.54	& 9.261	& 8.97 \\
   6	&12.42	&11.29	&10.680	&10.11 \\
   7	&11.85	&10.86	&10.330	& 9.81 \\
   8 	&15.08	&14.28	&13.803	&13.28 \\
\enddata
\\
a) Stars are identified in Figure \ref{fig:allsky}.\\
\label{table:allsky}
\end{deluxetable}

\begin{deluxetable}{lcl}
\tabletypesize{\small}
\tablewidth{0pt}
\tablecaption{{\rm The Star XO-1}}
\startdata
\hline
\hline
Parameter & Value & Reference\\
\hline
RA (J2000.0) & $ 16^h02^m11^s.84 $ & a,b \\
Dec (J2000.0) & +28\arcdeg10\arcmin10\arcsec.4 & a,b \\
$V, V_T$ & 11.19$\pm$0.03,11.323 & c,b\\
$B-V, B_T-V_T$ & 0.66$\pm$0.05,0.692 & c,b\\
$V-R_C$ & 0.38$\pm$0.04 & c\\
$R_C-I_C$ & 0.38$\pm$0.04,& c\\
$J$   & 9.939 & d\\
$J-H$ & 0.338 & d\\
$H-K$ & 0.074 & d\\
Spectral Type & \sptype & c \\
d & $200\pm20$ pc & c \\
$(\mu_\alpha,\mu_\delta)$ & ${\rm (-19.7 \pm 2.0, 15.0 \pm 1.9)~mas~yr^{-1}}$ & b \\
GSC 	& 02041-01657 & a \\
\enddata
\\
References:\\
a) SIMBAD\\
b) Tycho-2 Catalogue, Hog et al (2000) \\
c) this work \\
d) 2MASS, Skrutskie e al. (2006) \\
\label{table:star}
\end{deluxetable}
                                                                                            
\begin{deluxetable}{lcl}
\tabletypesize{\small}
\tablewidth{0pt}
\tablecaption{{\rm The Planet XO-1b}}
\startdata
\hline
\hline
Parameter & Value & Notes\\
\hline
$P $ 				& \vperiod$\pm$\eperiod\ d 		& \\
$t_c $	 			& \vjd$\pm$\ejd\ (HJD)			& \\
$K $ 				& \vrvK$\pm\ervK$\ \mps 			& \\
$R_{\rm p}/R_{\rm s} $ 		& (\vRp$\pm$0.04) $\times$ \Rjup/\Rsun 	& a\\
$a $ 				& \vap$\pm$\eap\ A.U. 			& b \\
$i $ 				& \vincl$\pm$\eincl\ deg		& b,c\\
$M_{\rm p} $ 			& \vMp$\pm$\eMp\ \Mjup		 	& b,d\\
$R_{\rm p} $ 			& \vRp$\pm$\eRp\ \Rjup			& a,b,c\\
\enddata
\\
Notes:\\
a) \Rjup\ = 71492 km, i.e. the equatorial radius of Jupiter\\
b) for $M_*$ = \vMs$\pm$\eMs\ \Msun \\
c) for $R_*$ = \vRs$\pm$\eRs\ \Rsun \\
d) \Mjup\ = 1.8988e27 kg 
\label{table:planet}
\end{deluxetable}

\begin{deluxetable}{cccc}
\tabletypesize{\small}
\tablewidth{0pt}
\tablecaption{{\rm Radial Velocity Shifts}}
\startdata
\hline
\hline
Julian Date & Radial Velocity &  Uncertainty & Telescope 	\\
            &  Shift [\mps] &  (1 $\sigma$) [\mps]  &	\\
\hline
    2453770.947  &    -99  &   63  & HJS 2.7-m \\
    2453771.945  &     93  &   70  & HJS \\
    2453773.924  &   -107  &   63  & HJS \\
    2453775.936  &    122  &   64  & HJS \\
\\
    2453779.941  &     87  &   15 & HET 11-m \\
    2453796.894  &     41  &   17 & HET \\
    2453799.884  &    112  &   14 & HET \\
    2453801.887  &   -111  &   15 & HET \\
    2453815.848  &    128  &   21 & HET \\
    2453824.815  &    -34  &   18 & HET \\
\enddata
\label{table:rv}
\end{deluxetable}

\begin{deluxetable}{ccc}
\tabletypesize{\small}
\tablewidth{0pt}
\tablecaption{{\rm Results of the SME Analysis}}
\startdata
\hline
\hline
Parameter & Mean &  Precision \\
          &      &  (1 $\sigma$)	\\
\hline
$T_{eff} [K]$	&5750 	&  13		\\
log$g$ [\cmpss]	&4.53  	&   0.065	\\
V~sin~$i$ [\kps] &1.11 	&   0.67	\\
\ [M/H]     	&0.058	&   0.040	\\
\ [Na/H]	&-0.051	&   0.041	\\
\ [Si/H]	&-0.043	&   0.038	\\
\ [Ti/H]	&-0.000	&   0.040	\\
\ [Fe/H]	&0.015 	&   0.040	\\
\ [Ni/H]	&-0.060	&   0.021	\\
\ [Si/Fe]   	&-0.058	&   0.056	\\
\enddata
\label{table:sme}
\end{deluxetable}

\begin{deluxetable}{lccc}
\tabletypesize{\small}
\tablewidth{0pt}
\tablecaption{{\rm Spectroscopically Derived Stellar parameters}}
\startdata
\hline
\hline
Parameter	&	@ 180 pc&	@ 200 pc&	@ 220 pc\\
\hline
   		&	0.98	&	0.97	&	0.97	\\
Mass [\Msun]   	&	1.01	&	1.00	&	1.00	\\
    		&	1.03	&	1.02	&	1.02	\\
    		&	    	&	    	&	    	\\
    		&	0.92	&	0.96	&	1.04	\\
Radius [\Rsun] 	&	0.95	&	1.00	&	1.09	\\
    		&	0.98	&	1.06	&	1.15	\\
    		&	    	&	    	&	    	\\
    		&	4.44	&	4.38	&	4.31	\\
Log(g) [\cmpss] &	4.49	&	4.43	&	4.35	\\
    		&	4.52	&	4.48	&	4.40	\\
    		&	    	&	    	&	    	\\
    		&	0.65	&	2.48	&	5.54	\\
Age [Gyr]     	&	2.30	&	4.74	&	6.93	\\
    		&	4.42	&	6.63	&	8.01	\\
\enddata
\\
For each parameter, the middle row is the maximum likelihood value, and the\\
values in the rows above and below span the 68\% likelihood of the probability\\
distributions (cf. Figure \protect{\ref{fig:stellarparam}}). The three columns correspond to three assumed\\
distances for XO-1.\\
\label{table:stellarparam}
\end{deluxetable}

\end{document}